\documentclass[12pt,preprint]{aastex}

\newcommand{\beq}{\begin{equation}}
\newcommand{\eeq}{\end{equation}}

\def\3he{$^3$He\,}
\def\he4{$^4$He\,}

\shorttitle{Impulsive Thermal X-rays}
\shortauthors{Liu et al.}

\begin{document}

\title{Impulsive Thermal X-ray Emission from a Low-lying Coronal Loop}

\author{
Siming Liu\altaffilmark{1}, Youping Li\altaffilmark{1}, and Lyndsay Fletcher\altaffilmark{2}
}
\altaffiltext{1}{Key Laboratory of Dark Matter and Space Astronomy, Purple Mountain Observatory, Chinese Academy of Sciences,
Nanjing, 210008, P. R. China; liusm@pmo.ac.cn}
\altaffiltext{2}{Department of Physics and Astronomy, University of
Glasgow, Glasgow, G12 8QQ, Scotland}

\begin{abstract}
Understanding the relationship among different emission components plays an essential role in the study of particle acceleration and energy conversion in solar flares.
In flares where gradual and impulsive emission components can be readily identified the impulsive emission has been attributed to non-thermal particles. We carry out detailed analysis of H$\alpha$ and X-ray observations of a GOES class B microflare loop on the solar disk. The impulsive hard X-ray emission, however, is found to be consistent with a hot, quasi-thermal origin, and there is little evidence of emission from chromospheric footpoints, which challenges conventional models of flares and reveals a class of microflares associated with dense loops.
H$\alpha$ observations indicate that the loop lies very low in the solar corona or even in the chromosphere and both emission and absorption materials evolve during the flare. The enhanced H$\alpha$ emission may very well originate from the photosphere when the low-lying flare loop heats up the underlying chromosphere and reduces the corresponding H$\alpha$ opacity.
These observations may be compared with detailed modeling of flare loops with the internal kink instability, where the mode remains confined in space without apparent change in the global field shape, to uncover the underlying physical processes and to probe the structure of solar atmosphere.
\end{abstract}


\section{INTRODUCTION}
\label{intro}
Most observational studies of particle acceleration in solar flares focus on large events that usually have good observational coverage with an energy release up to $10^{32-33}$ ergs \citep[e.g.,][]{m94, l03}. Given the continuous distribution of many flare properties, the energy release in small flares have been assumed to be similar to that in large flares and the inferred physical processes from observations of large flares should operate in a similar way in small flares \citep{v02a}. The observed similarities between large and small flares have also been used to argue for common physical processes in flares of different magnitude. Even most transient X-ray brightenings discovered with the Soft X-ray Telescope on Yohkoh satellite with an energy release of $10^{25-29}$ ergs \citep{s92} have been identified as microflares or mini-flares \citep{n97,n99}. These studies emphasize the common characteristics of transient energy release in the solar corona \citep{b01}.

On the other hand, there is also evidence that large flares behave differently from small ones \citep{m99, v02b}. Flares originate from the complex magnetic environment of the solar corona, that allows some unique characteristics for each flare. Processes identified from observations of one flare may not be significant in another. Flares, as identified with some common observational characteristics, certainly share some common physical processes, e.g. the relevant radiative processes, which typically do not depend on the flare magnitude and the magnetic field geometry. However, the energy release and particle acceleration processes may carry some intrinsic scales and magnetic field structure. Their importance in a given flare will depend on the flare magnitude and the topological evolution of the magnetic field. Indeed, many flare models have been proposed based mostly on the magnetic field topology
\footnote{http://solarmuri.ssl.berkeley.edu/$\sim$hhudson/cartoons/}.
Given the complex flare environment, the continuous distribution of many flare characteristics may just suggest a continuous distribution of the magnetic field structure and/or a lack of distinct scales in these phenomena instead of common energy release and particle acceleration mechanisms for all flares \citep{l93}. Detailed studies of individual flares, which usually reveal peculiarities of specific flares, are also essential for better understanding of the flare phenomena \citep[e.g.,][]{s10, b12}. In this paper, we carried out detailed analysis of multi-wavelength observations of a low-lying flare loop on the solar disk that has several unique characteristics.

Two emission components can be readily identified from X-ray observations of many flares: impulsive high-energy emission and gradual low-energy emission, which have been associated with non-thermal and thermal particles, respectively\citep{g11}. In light of the Neupert effect \citep{n68}, \citet{b71} first proposed the classical thick-target flare model \citep{b73}, which has been explored extensively by later studies \citep{v05a}. In this paper, we carry out detailed analysis of X-ray (with the GOES and Ramaty High Energy Solar Spectroscopic Imager: RHESSI), EUV (with the EIT), and H$\alpha$ (with the Big Bear Solar Observatory: BBSO) observations of a small GOES class B flare with distinct impulsive and gradual hard X-ray emission components. Our results show that the impulsive emission likely has a quasi-thermal origin and there is no evidence of emission from chromospheric footpoints of the loop, which contradicts the conventional interpretation but is consistent with the high-temperature turbulence-current sheet (HTTCS) model \citep{lm93, s97, s98}. The flare also has a relatively stable global linear structure, which may be caused by an internal kink instability proposed for active region brightenings \citep{h07}. Observations of similar flares with better observational coverage and corresponding numerical simulations can be used to probe the structure of similar loops and the solar atmosphere.
Analysis of observations of the flare is presented in \S\ \ref{obs}. The implication of these results are discussed in \S\ \ref{dis}, where we also draw our conclusions.

\section{OBSERVATIONS AND DATA ANALYSIS}
\label{obs}

The flare studied here occurred on 26 June 2002. It is selected due to its relatively simple light curves (first panels of Fig. \ref{fig1}). The GOES soft X-ray fluxes increase monotonically from $\sim$ 18:47 to $\sim$ 18:51. The decay phase afterward lasts for more than 10 minutes (right panel of Fig. \ref{fig3}). Based on the peak flux in the 1-8 \AA\ waveband, the flare is classified between B7 and B8. A quiet period before 18:46 can be selected as the background. The background subtracted peak flux in the 1-8 \AA\ waveband is about $4.3\times 10^{-7}$ W m$^{-2}$.

RHESSI observations reveal a prominent pulse in the 12-25 keV energy band (Fig. \ref{fig1} right first panel), which indicates a distinct impulsive emission component. The pulse lasts for $\sim$ 40 seconds, followed by a gradually varying emission component. We identify this pulse as the impulsive phase. There is no evidence of emission above 25 keV and the light curves below 12 keV are similar to those GOES light curves. The time-derivative of the GOES fluxes are almost constant from 18:47:40 to 18:51:00 and do not have distinct feature correlated with the hard X-ray (HXR) pulse (second panels of Fig. \ref{fig1}), in contrast to the Neupert effect usually observed in large flares \citep{n68}.

\subsection{X-ray Spectral Evolution}
\label{spec0}

\begin{figure}[ht]
\begin{center}
\includegraphics[width=14cm]{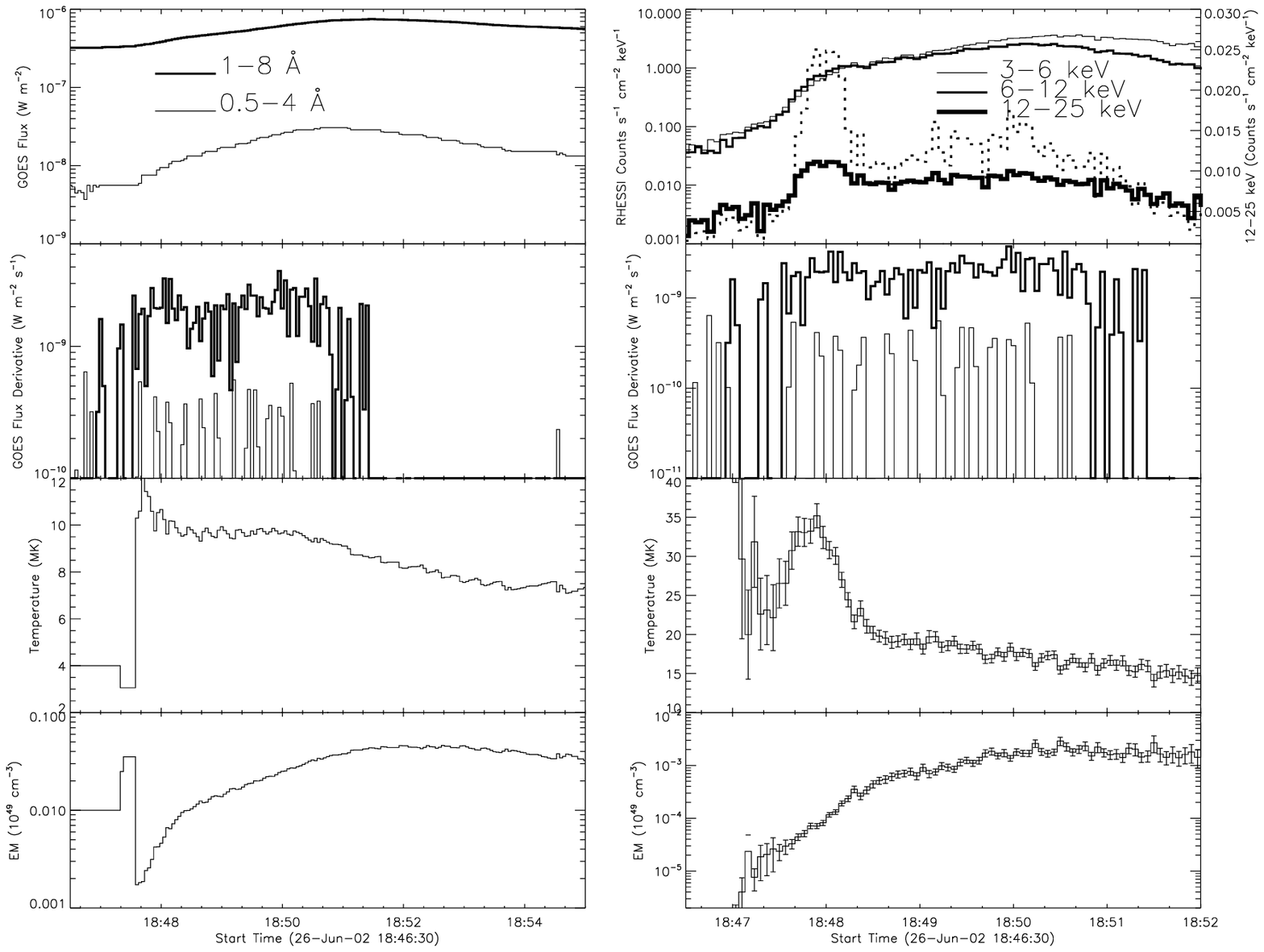}
\caption{
 Summary of spectral fits of GOES (left) and RHESSI (right except for the second panel that is the same as the second panel on the left)
 observations. Note the time interval on the left covers that on the right. The top panels give the GOES fluxes and RHESSI count rates.
 The thick and thin lines in the left panel correspond to the 1-8 \AA\
 and 0.5-4 \AA\ wavebands, respectively.
 In the right panel, the 12-25 keV count rate is also shown in linear scale indicated on the right axis with the dotted line to demonstrate the impulsive component clearly. The impulsive phase is defined by the prominent pulse from $\sim$ 18:47:40 to $\sim$ 18:48:20.
 The second panels give the logarithmic plots of the time-derivatives of the GOES fluxes with the same line type as the top panel on the left. Missing data intervals correspond to decreases with time in the corresponding fluxes.
 %
 The third and fourth panels are for the temperature and EM, respectively. The Coronal model with Chianti version 6.0.1 is used for the background subtracted GOES data. An isothermal model with full Chianti version 5.2 is used to fit the RHESSI spectrum above 7 keV.
}
\label{fig1}
\end{center}
\end{figure}

The Coronal model with Chianti version 6.0.1 is used to fit the background subtracted GOES fluxes. The left panel of Figure \ref{fig1} shows the evolution of the temperature (third panel) and emission measure (fourth panel). Reliable spectral parameters can be obtained after $\sim$ 18:47:40. The temperature has a nearly constant value of 10 MK from $\sim$ 18:47:40 to $\sim$ 18:50:20 and decreases gradually afterward. The emission measure (EM) appears to experience two phases of growth. The growth rate in the impulsive phase from 18:47:40 to 18:48:20 is much higher than that afterward. The EM reaches a peak value of $\sim 5\times 10^{47}$ cm$^{-3}$ at $\sim$ 18:52:00 and decays gradually thereafter.
The total radiation power of this GOES component peaks at $\sim 4\times 10^{25}$ erg s$^{-1}$ near 18:52:30 and the total radiative energy loss during the flare is on the order of $10^{28}$ ergs.

\begin{figure}[ht]
\begin{center}
\includegraphics[width=14cm]{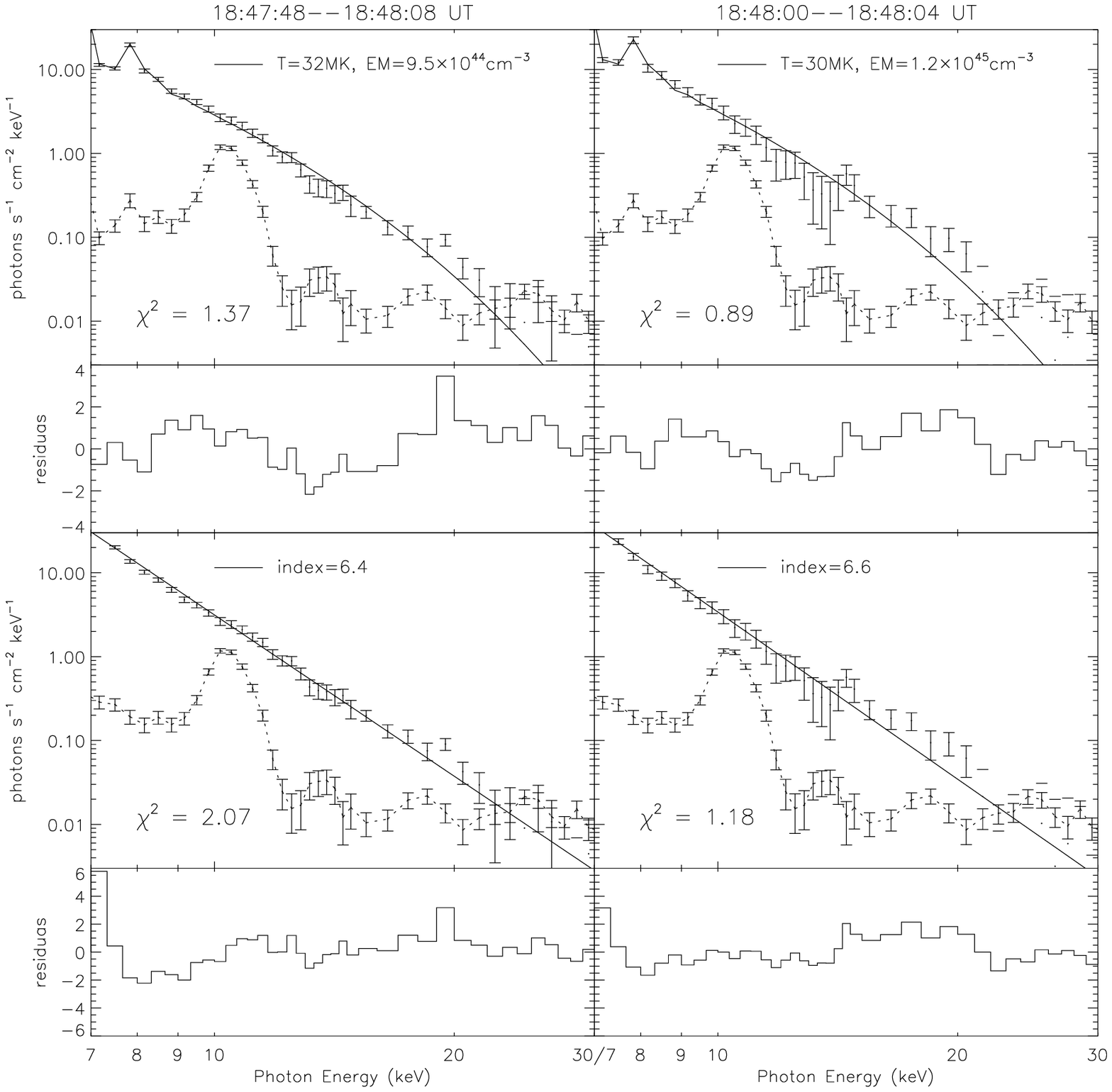}
\vspace{-0mm}
\caption{
Spectral fits to a 20 second (left) and a 4 second (right) interval near
 the HXR peak. Only the energy range from 7 to 30 keV and front segment of detectors 1, 3, 6 are considered. An isothermal model (top) and a single power-law model (bottom) are assumed for the spatially integrated emission.
  In both the top and bottom rows, the model parameters and the reduced $\chi^2$ are indicated in the upper panels that show the photon spectra of the observation, the model (solid line), and the background (connected by a dashed line). The lower panels show the normalized residuals. For the power-law model, the residual is highest at the low energy end due to the absence of emission lines in the model spectrum. Note that for a given time interval different photon spectra can be derived from the observed count rate spectrum for different emission models. The background photon spectrum however is the same for all intervals and models.
}
\label{fig2}
\end{center}
\end{figure}

For every 4 second interval, we fit the RHESSI spectrum with an isothermal model. Chianti version 5.2 is used to derive the full isothermal model spectrum.
A quiet period before 18:46:00 is selected as the background and the RHESSI background spectrum is shown as data points connected by a dashed line in Figure \ref{fig2}. The spectral fit is carried out for individual detectors and results from the front segment of detectors 1, 3, 6 agree with each other above 7 keV \footnote{We find that residuals of spectral fit for other detectors have some systematic patterns and do not converge as well as detectors 1, 3, and 6.}. We therefore only consider the combined spectrum from detectors 1, 3, and 6 above 7 keV \citep{g11}. The third and fourth panels of the right panel of Figure \ref{fig1} show the evolution of the temperature and EM. During the HXR pulse, the temperature reaches a peak value of 35 MK near 18:48, which is consistent with the superhot component in the HTTCS model \citep{s98} and decreases to $\sim$ 25 MK within 25 seconds. The temperature decreases almost linearly after the HXR pulse. The EM increases by almost one order of magnitude during the HXR pulse and its growth becomes more gradual afterward, which agrees with the evolution of the EM derived from the GOES data. The EM determined from the RHESSI data reaches a peak value of about $2\times 10^{46}$ cm$^{-3}$ near 18:50, which is more than one order of magnitude lower than the peak value of the EM obtained with GOES observations.

The top row of Figure \ref{fig2} shows two examples of the spectral fit near the HXR peak with the isothermal model. HXR pulses are usually attributed to bremsstrahlung emission of electrons with a power-law distribution. The bottom row of the figure shows single power-law model fits for the same two time intervals. The model gives similar values of the reduced $\chi^2$ as the isothermal model. However, the photon spectrum is very soft with an index always greater than 6. We also compare the residuals of isothermal and single power law spectral fits at different time intervals. No systematic pattern is observed in either case except that the normalized residual for the power-law model is highest near the low energy end of 7 keV due to the absence of emission lines in the model (lower panels in the bottom row of Fig. \ref{fig2}). The isothermal and single power law models give equally acceptable fits to the spectra. Through the spectral fit alone, we therefore cannot distinguish a thermal or non-thermal origin of the HXR pulse. As we will show below, other observations of this flare favor a thermal explanation for the HXR pulse.

\subsection{X-ray, H$\alpha$, and EUV Images}
\label{ha}

\begin{figure}[ht]
\begin{center}
\includegraphics[width=8cm]{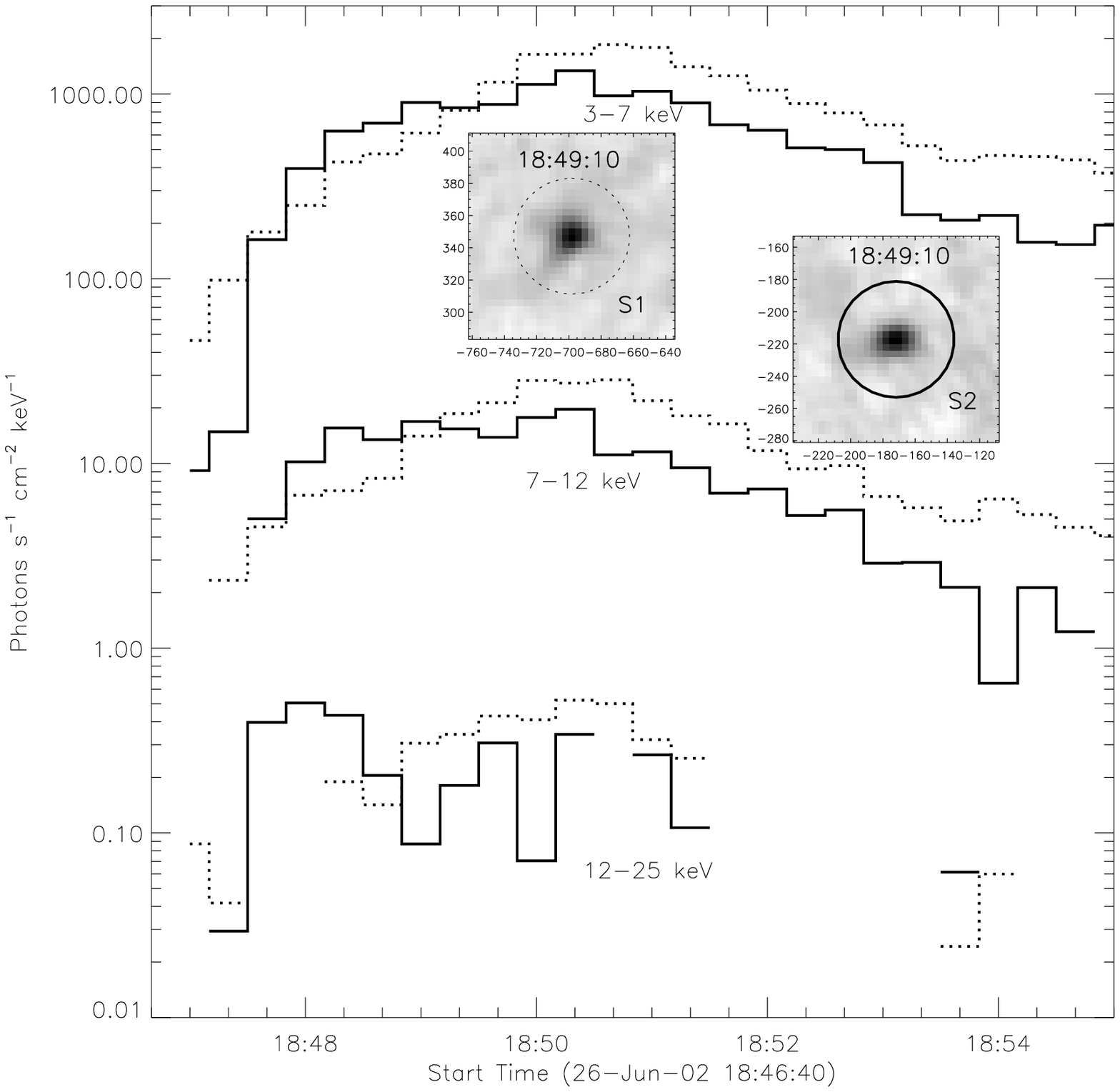}
\includegraphics[width=7.8cm]{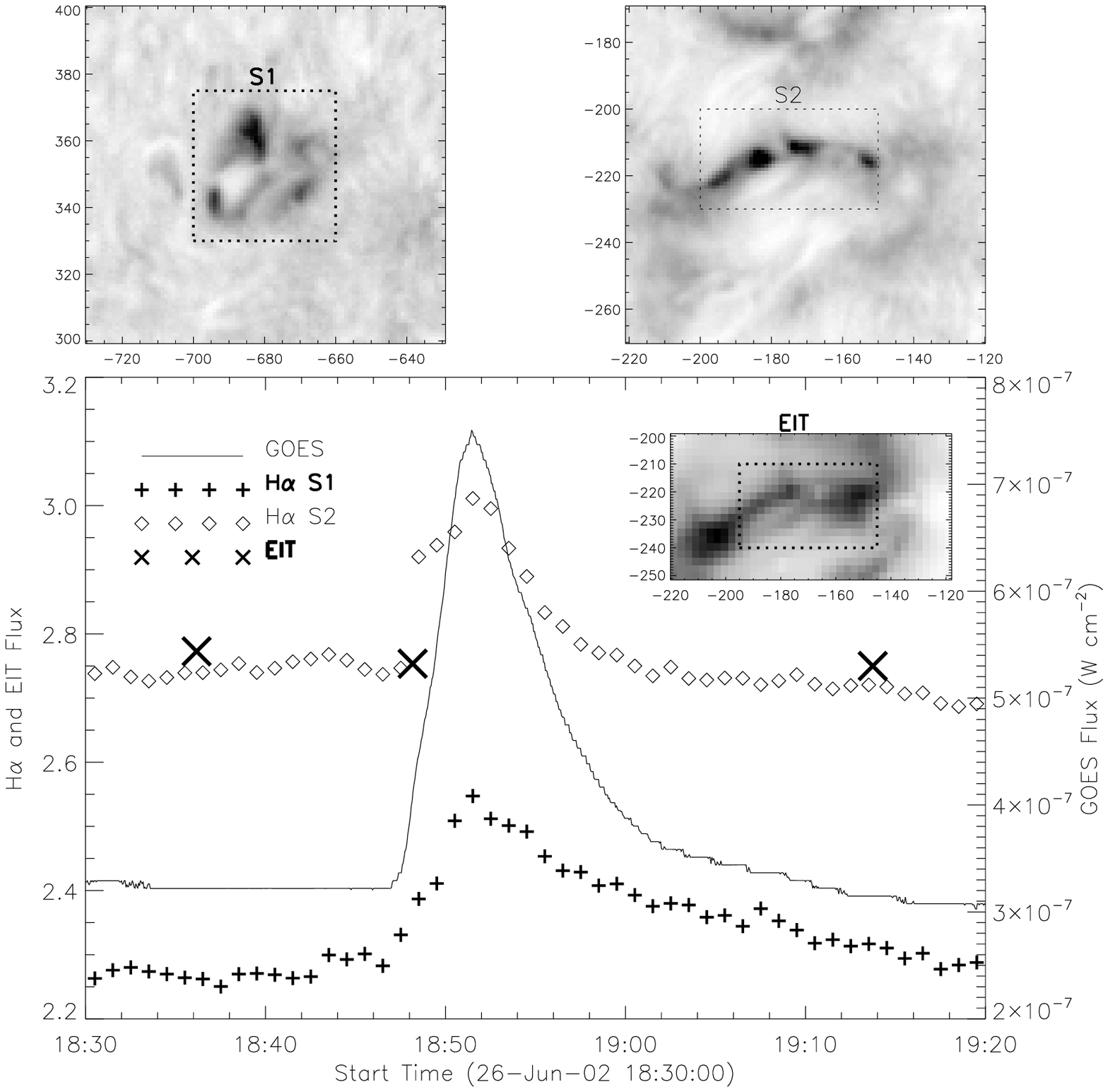}
\vspace{-0mm}
\caption{
Light curves of two flares occurring within a few seconds but spatially separated by more than $6\times 10^{10}$ cm on the solar surface. The left panel shows the RHESSI light curves. Images are made for every 20 second interval with the CLEAN algorithm with the front segment of all detectors except for detector 2 considered. The inserted images are for the time interval starting at 18:49:10. The circles indicate the regions where the integrated X-ray fluxes are extracted. The dotted and solid lines are for S1 in the northern hemisphere and S2 in the southern hemisphere, respectively. The HXR pulse in the 12-25 keV band is associated with S2. The missing data points correspond to images with poor quality so that no reliable flux can be extracted from the chosen regions.
The right panel shows H$\alpha$ and EIT observations of these two flares together with the GOES 1-8 \AA\ light curve. The dashed boxes in the inserted images indicate the regions where the fluxes are extracted. The H$\alpha$ images of the top two panels were taken at 18:48:31. The lower EIT image was taken near 18:48. The three crosses are for EIT 195 \AA\ fluxes of S2 with an arbitrary units. There is no evidence of enhanced EUV emission at the peak of the HXR pulse at 18:48.
}
\label{fig3}
\end{center}
\end{figure}

A preliminary study of the RHESSI X-ray image of this flare reveals two flares separated by about $52^\circ$ on the solar surface. The left panel of Figure \ref{fig3} shows the light curves of these two flares in three X-ray bands. It is evident that the HXR pulse in the 12-25 keV energy band is associated with the flare in the southern hemisphere (S2) with position ($102^\circ$, $-12^\circ$) in heliographic coordinates with $x$-axis along the line of sight pointing toward the Earth and $z$-axis to the north. The X-ray fluxes of S2 peak earlier at higher energies similar to other impulsive flares, and the X-ray fluxes do not change significantly from 18:48 to 18:51. The other source (S1: the default source identified by the RHESSI software) is located in the northern hemisphere with coordinate ($68^\circ$, $-52^\circ$) and has more gradual time evolution and better correlated fluxes among different X-ray bands. These two flares appear to start within a few seconds of one another. For a spatial separation of $6.3\times 10^{10}$ cm on the solar surface, they are not likely causally connected. Sympathetic flares have been studied extensively before \citep[e.g.,][]{f76, n85}. These two flares are both GOES B class small flares with no apparent agent linking the two flaring active regions. The fact that they occurred almost simultaneously with comparable X-ray fluxes at the peak appears to be a coincidence. Given the small magnitude of these flares, they are also transient brightenings in active regions, which frequently occur simultaneously without obvious physical connections \citep{s92}.

H$\alpha$ images of the Sun are routinely taken at the BBSO with a cadence of 1 minutes. The right panel of Figure \ref{fig3} shows the H$\alpha$ light curves of these two flares. It is interesting to note that the H$\alpha$ fluxes of both flares peak at the same time of $\sim$ 18:51:30 as the GOES 1-8 \AA\ flux.
The H$\alpha$ flux of S2 increases dramatically from 18:47:31 to 18:48:31, which appears to be correlated with the HXR pulse. Both the X-ray and H$\alpha$ light curves of S1 rise more gradually than those of S2. The sharp rise of the GOES EM during the HXR pulse as shown in Figure \ref{fig1} therefore is likely caused by S2. As with the RHESSI observations, the H$\alpha$ fluxes of these two flares have similar amplitudes.
Although the box designed to extract flux from S1 is larger than that for S2, its H$\alpha$ flux is lower than that of S2 due to the limb darkening effect. However, the amplitude of the H$\alpha$ flux increases during these flares are comparable.

S1 has complex structures in both H$\alpha$ and X-ray images and its fluxes in different wavelengths show well correlated gradual evolution as seen in gradual flares. S2 has a distinct impulsive emission component and behaves like an impulsive flare with higher energy emission peaking at an earlier time. We will focus on analyzing S2 in the following. The EIT 195 \AA\ flux from the same source region as S2 is indicated by the three crosses in the right panel of Figure \ref{fig3}. The second data point corresponds to the peak of the HXR pulse at $\sim $ 18:48. There is no evidence of EUV emission associated with the flare S2. We also search the data archive. There are no TRACE observations of this flare.

\begin{figure}[ht]
\begin{center}
\includegraphics[width=15cm]{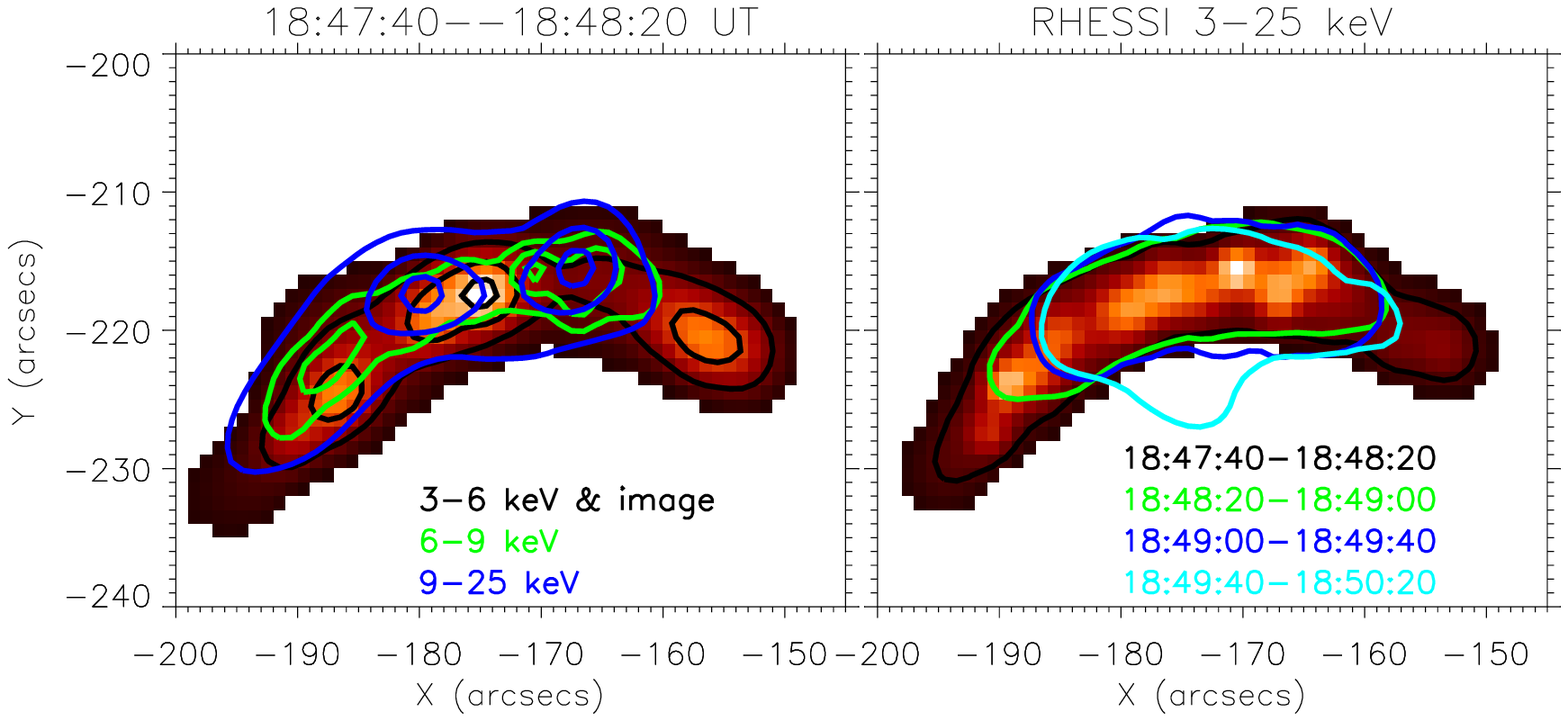}
\vspace{-0mm}
\caption{
{\it Left:} Image of the flare loop in the impulsive phase from 18:47:40 to 18:48:20 at 3-6 keV. The Pixon algorithm with the front segment of detectors 1, 3, 4, 5, 6, 8, and 9 is used to resolve fine structures in these images. The black overlaid contour levels indicate 30, 60,
 and 90\% of the peak intensity. The green and blue contours
 are of the same contour levels but for the 6-9 keV and 9-25 keV energy
 bands, respectively. {\it Right:} The impulsive phase (from 18:47:40 to 18:48:20) image at 3-25 keV, which is consistent with the {\it left} panel. The longest contour indicates 20\% of the peak value. The other contours show the same
 relative contour level but for a continuous series of 40 second intervals after the
 impulsive phase.
}
\label{fig4}
\end{center}
\end{figure}

{\bf X-ray Images:} To study the fine X-ray structure of the flare S2, we use the front segment of detectors 1, 3, 4, 5, 6, 8, and 9 and the Pixon algorithm to reconstruct the source images. Although the Pixon algorithm is much more time consuming than the CLEAN algorithm, it can achieve better spatial resolution and dynamical range and is needed to study fine structures \citep{a03}. The left panel of Figure \ref{fig4} shows the images of the HXR pulse from 18:47:40 to 18:48:20 in the 3-6, 6-9, and 9-25 keV bands. Although all X-ray emissions align along an arc structure, there are small scale structures, which are not well correlated across different energy channels. The right panel of Figure \ref{fig4} shows the time evolution of the X-ray source. The image is for the HXR pulse and the whole 3-25 keV energy band. The X-ray source has significant clumps along the arc structure and is longest in the impulsive phase.
As the flare evolves, it becomes more and more compact and the contraction is always along the loop.

These complex source structures, in combination with the fact that the other flare S1 with a similar magnitude occurs simultaneously with the flare loop, show clearly that the interpretation of the spectral fit results can be highly ambiguous. Although both an isothermal and a single power law model give reasonable fits to the spatially integrated spectra, the fact that these two flares are independent and have complex energy-dependent structures strongly rejects literal interpretation of the spectral fit results. A physical isothermal source will have well correlated structure for X-ray images in different energy bands. Injection of a power-law electron population into a simple loop will also produce well defined energy dependent source structure \citep{a02}. The complex energy dependent source structure therefore shows that the X-ray loop neither corresponds to an isothermal source, nor can it be explained with simple models of injections of an electron population with a single power law distribution into the flare region.
However in comparison with the single power-law model, the isothermal model is favored due to the very high values of the spectral index of the power law model.
The combination of imaging and spectroscopic observations of RHESSI therefore plays essential roles in uncovering the underlying physical processes.

Given the complex X-ray structure, the X-ray arc may be associated with a loop with twisted magnetic field lines. The flare can be triggered by reconnections within the loop \citep{h07,p10, b12}. Loops can carry plasmas with different temperatures in different strands due to suppression of cross-field thermal conduction by the strong magnetic field. There is also ample observational evidence for multi-temperature plasmas in a flare loop \citep{f03, s10}.  In combination with GOES observations and H$\alpha$ observations studied below, a multi-temperature loop gives the most reasonable explanation of the flare \footnote{It may be tempting to do an imaging spectroscopy study of these two flares. Given the low magnitude of these flares and the ambiguity of the fitting to the spatially integrated spectra, such a study is not expected to give further insights to the nature of these flares.}.

There is no evidence of emission from the chromospheric footpoints of the loop implying a thin target origin of the X-ray emission. The length of the X-ray loop is about 60$^{\prime\prime}$ and the width is about 10$^{\prime\prime}$, the corresponding volume is on the order of 10$^{27}$ cm$^{3}$. From the peak value of $\sim 5\times 10^{47}$ cm$^{-3}$ of the EM obtained from GOES observations and assuming a volume filling factor of 1, we infer a density on the order of $10^{10}$cm$^{-3}$, which is typical for the coronal source of a flare and is sufficient to confine low energy ($<25$ keV) electrons in the coronal part of the loop \citep{g11}. The total thermal energy content of the emitting plasma at the peak time of the GOES EM is on the order of $10^{28}$ ergs consistent with transient brightenings in active regions and the total radiative energy of the flare as inferred from GOES observations. The density of electrons producing the RHESSI source is a factor of a few lower than that inferred from GOES observations due to the lower values of the EM inferred from RHESSI spectral fits, which may suggest a lower filling factor of the HXR emitting plasma. Since a high density is needed to confine tens of keV electrons in the corona part of the loop during the rise phase, the increase of the EM in the rise phase is likely caused by an increase of the volume filling factor of the emitting plasmas instead of an increase of the density due to evaporation from the chromosphere footpoints, which is also consistent with the multi-strand interpretation of the flare loop inferred above.

\begin{figure}[ht]
\begin{center}
\includegraphics[width=8cm]{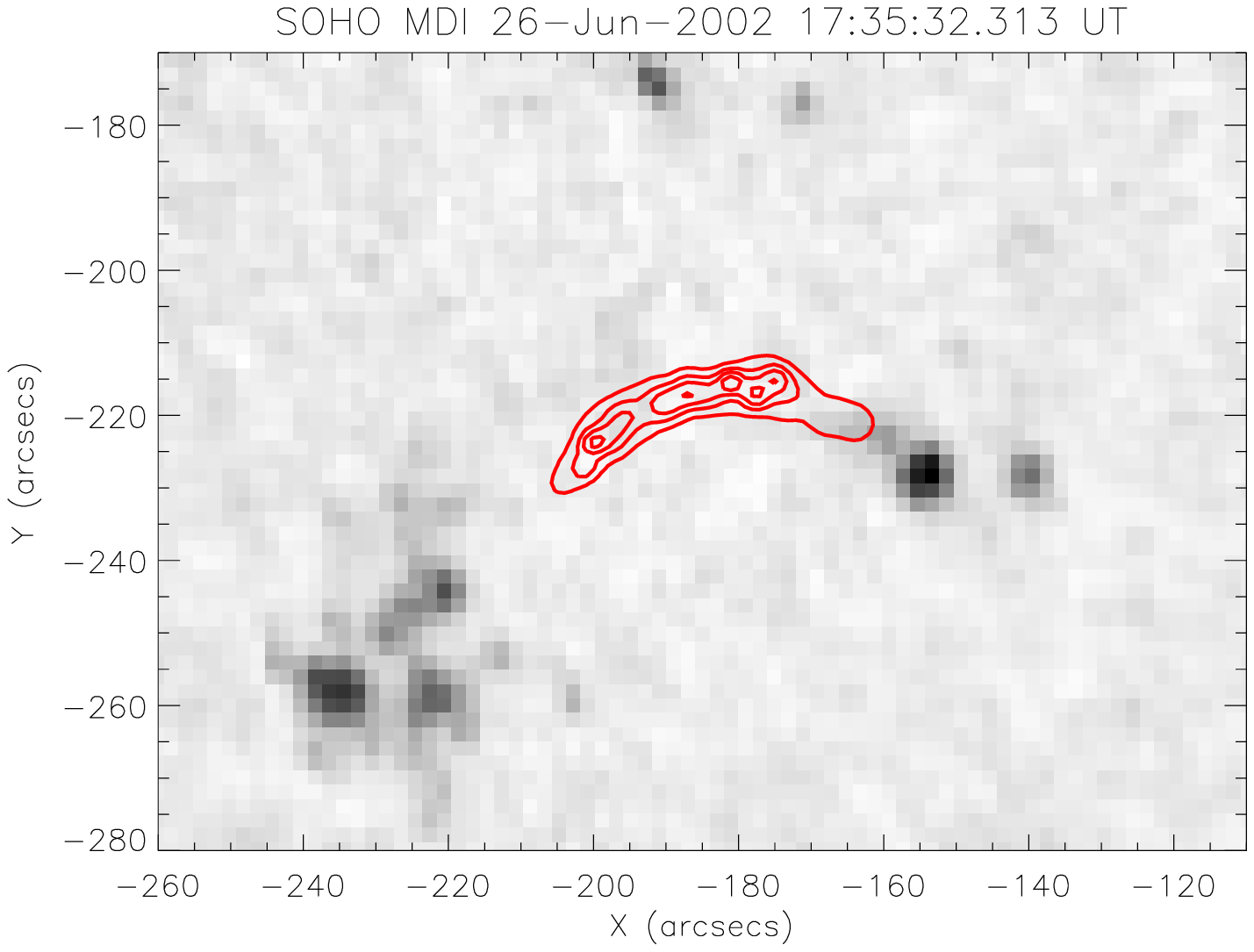}
\includegraphics[width=8cm]{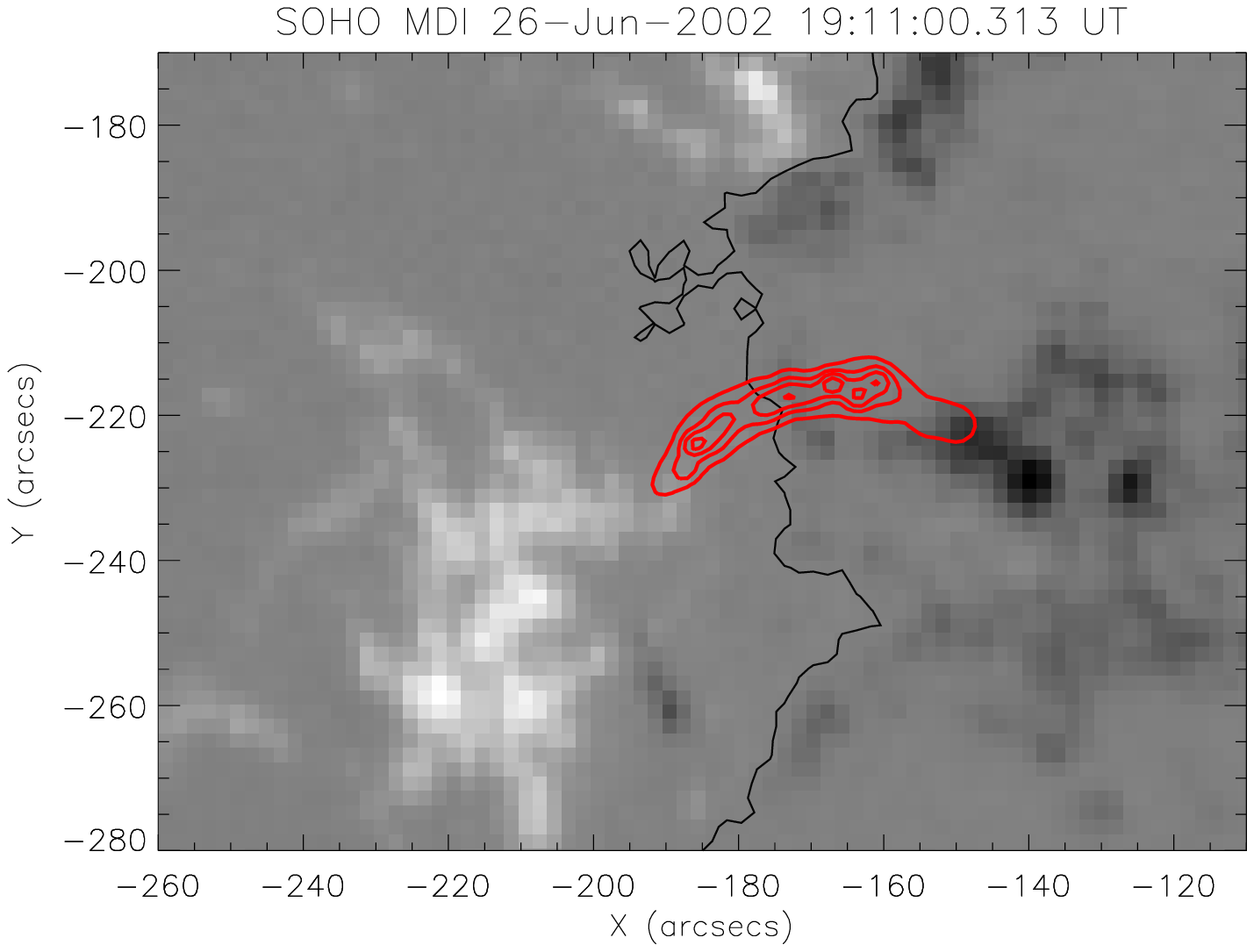}
\vspace{-0mm}
\caption{
{\it Left:} 20, 40, 60, 80\% contours of the impulsive phase (from 18:47:40 to 18:48:20) X-ray image at 3-25 keV (of the right panel of Figure \ref{fig4}) overlaid on an MDI continuum image taken at 17:35:32. The position of the RHESSI contours has been rotated to account for the effect of solar rotation. The flare loop points toward two dark regions. {\it Right:} Same as the left panel but for an MDI magnetogram taken at 19:11:00. The loop points toward two regions with opposite polarity. The thin line indicates the dominant line of magnetic polarity inversion.
}
\label{fig5}
\end{center}
\end{figure}

Figure \ref{fig5} shows the overlay of RHESSI contours during the HXR pulse from 18:47:40 to 18:48:20 with the MDI continuum (left) and magnetogram (right) images. Although these MDI images are not taken at the same time as the RHESSI image, the photospheric structures revealed by these images are stable and do not change significantly within a few hours. It is evident that the loop connects two regions with opposite magnetic polarities and the dominant line of magnetic polarity inversion is almost perpendicular to the arc of the loop. The magnetic field lines of the loop therefore likely originate from two footpoints of the loop instead of two ribbon structures for large flares.

\begin{figure}[ht]
\begin{center}
\includegraphics[width=8cm]{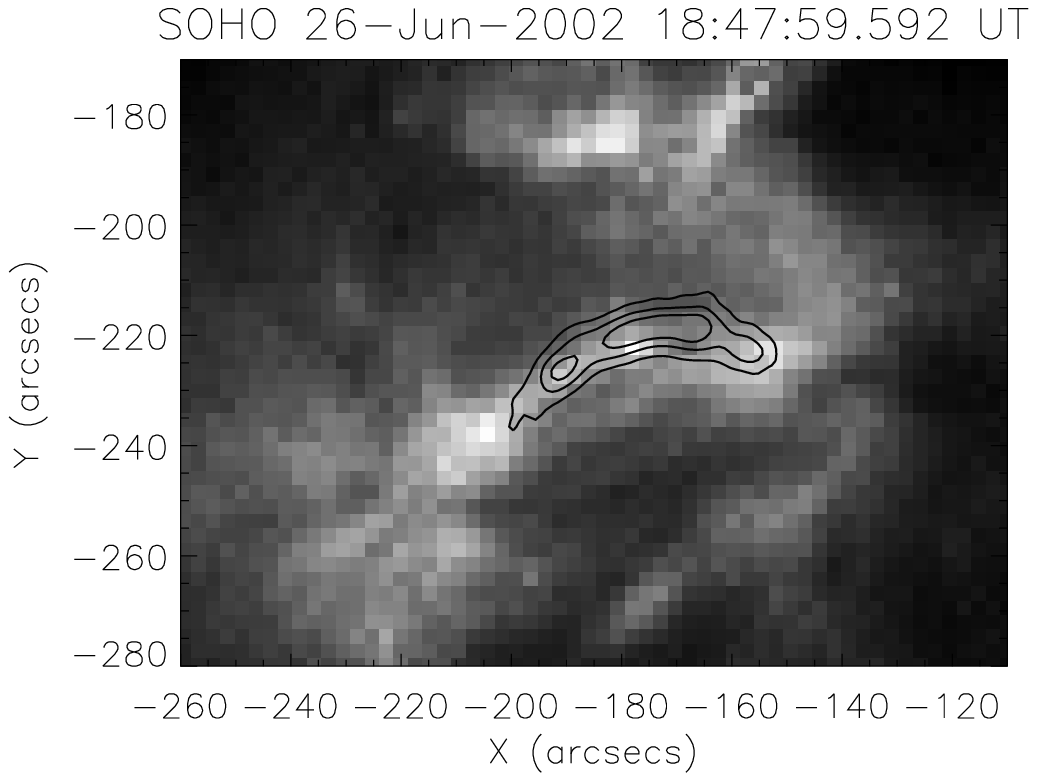}
\includegraphics[width=8cm]{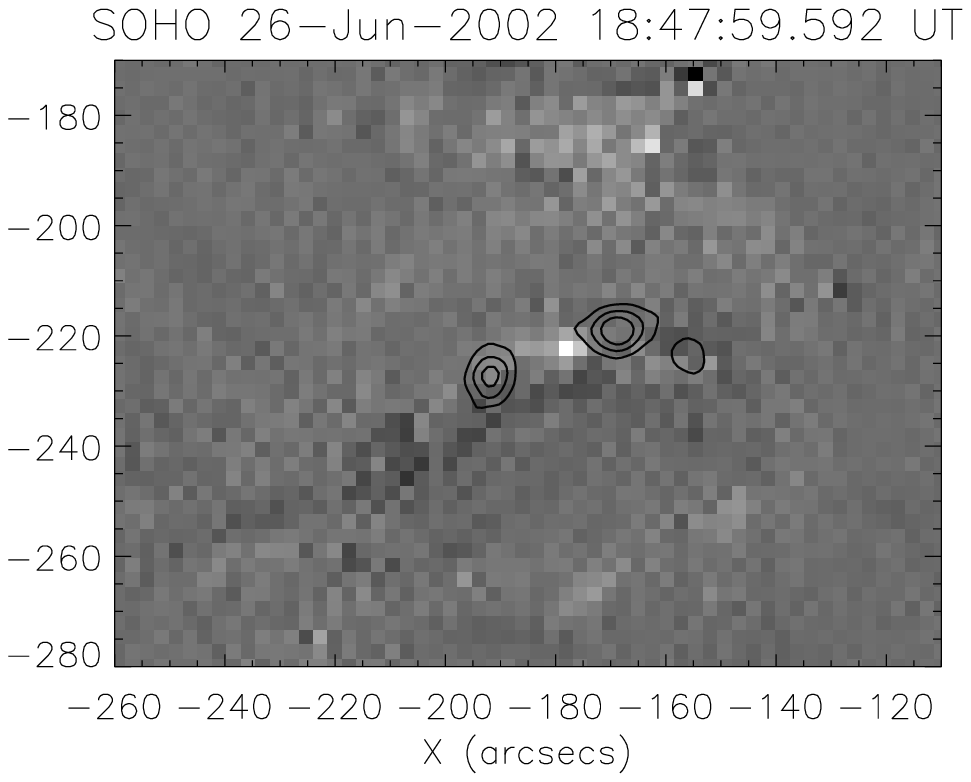}
\vspace{-0mm}
\caption{
{\it Left:} Same as Figure \ref{fig5} but for an EIT 195 \AA\ image taken at 18:48:00 and the X-ray contours are of 40, 60, 80\% of the peak intensity of a RHESSI image made with the CLEAN algorithm for the 3-25 keV range and for the HXR pulse from 18:47:40 to 18:48:20. The front segment of detectors 1, 3, 4, 5, 6, 7, 8, 9 are used. A bright EIT feature (presumably a loop) is well aligned with the X-ray loop.
 {\it Right:} The difference map between EIT images taken at 18:36 and 18:48. The contours are of 40, 60, 80\% of the peak intensity of a similar RHESSI image made with the CLEAN algorithm but for a period from 18:47:01 to 18:48:01 right before the EIT exposure time.  There are no new EIT features associated with the X-ray sources.
}
\label{fig6}
\end{center}
\end{figure}

{\bf EUV Images:} The SOHO EIT takes a full Sun image in the EUV band with a cadence of $\sim$ 12 minutes. These EUV observations reveal hot coronal plasmas especially those associated with coronal loops. Most EIT observations near the flare period were taken at 195 \AA. The left panel of Figure \ref{fig6} shows an EIT image taken at the peak of the HXR pulse. The bright coronal features revealed here are relatively stable. We made difference maps of EIT images of the active region of S2 taken at two neighboring times and plotted the occurrence frequency distribution of the flux variations. The distribution is consistent with a Gaussian except for a few pixels, which represent statistically significant variations.
We find that fluctuations near the bright EUV features are relatively higher and statistically significant (right panel of Fig. \ref{fig6}).

The flare X-ray loop is well aligned with a bright EUV loop (left panel of Fig.\ref{fig6}). However, no particular new EUV feature emerges at the peak of the HXR pulse. The right panel of Figure \ref{fig6} shows the 195 \AA\ difference map between 18:36 and 18:48. The contours are for X-ray observations from 18:47:01 to 18:48:01 right before the second EIT observation. Three X-ray sources can be readily identified. A bright EIT spot, whose amplitude is comparable to variations in the difference maps taken before the flare and in other EUV bright regions, is located between the left two X-ray sources. A post flare observation at 19:13:45, when the X-ray and H$\alpha$ fluxes have decreased to the pre-flare levels (see the right panel of \ref{fig3}), does not reveal any new features associated with the flare loop either. As we will show below, H$\alpha$ observations suggest that the flare loop has a very low altitude. The bright EUV loops are hot and should be high in the corona. They therefore may not participate in the flare process directly. Nevertheless the absence of EUV emission at the HXR pulse remains puzzling given the high temperature ($>20$ MK) inferred from the RHESSI spectral fit (right panel of Fig. \ref{fig1}.

{\bf H$\alpha$ Images:} The Big Bear Solar Observatory takes full Sun images at H$\alpha$ with a cadence of 1 minute and covers the period of the flare loop. To calibrate these images with RHESSI observations, we first use the solar limb to determine the disk center for each images, which can be identified with the disk center of the RHESSI images. The H$\alpha$ image is then cross-correlated with an SOHO MDI continuum image near a sunspot to the south-west to determine the displacement in the position angle. This displacement angle does not change dramatically from image to image. We therefore fix it to a mean value.

\begin{figure}[ht]
\begin{center}
\includegraphics[width=8cm]{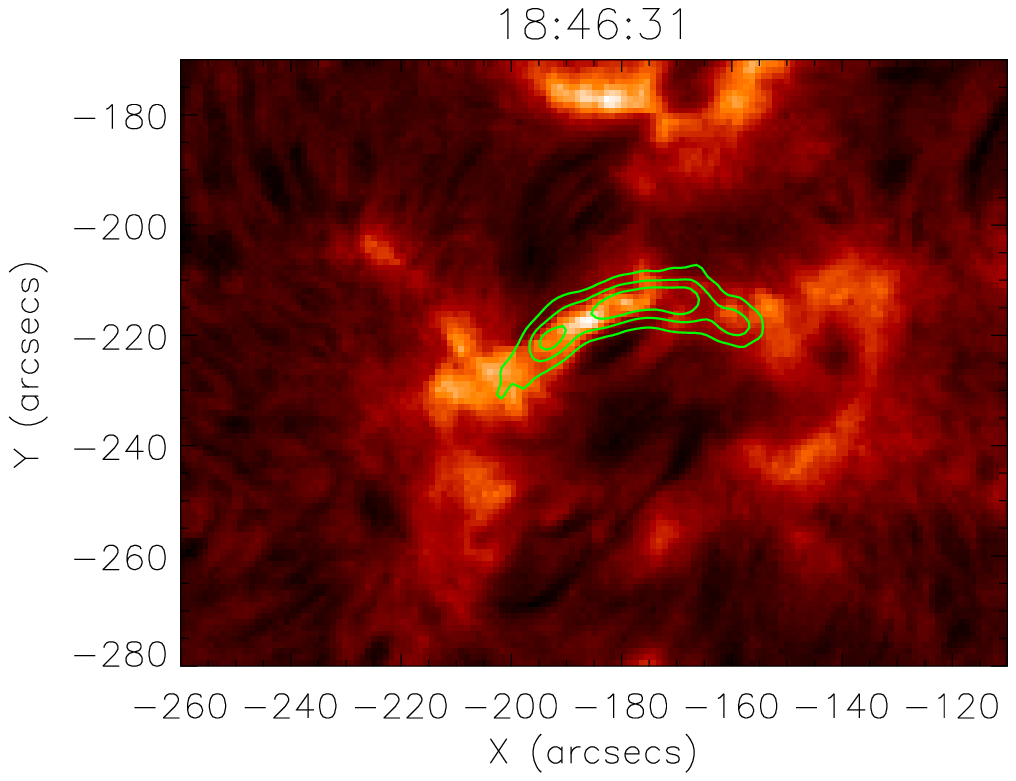}
\includegraphics[width=8cm]{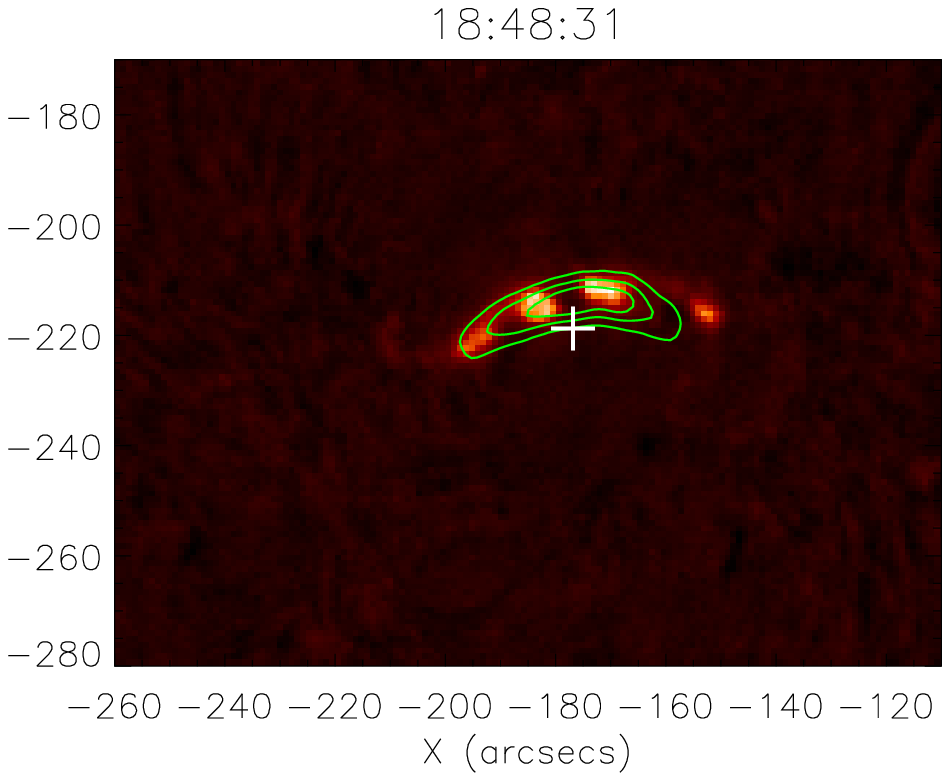}
\vspace{-0mm}
\caption{
{\it Left:} Same as the left panel of Figure \ref{fig6} but for an H$\alpha$ image taken at 18:46:31. The X-ray loop aligns with a bright H$\alpha$ arc.
{\it Right:} The difference map between H$\alpha$ images taken at 18:46:31 and 18:48:31. The X-ray contours are for a period from 18:48:01 to 18:49:01. Note that the absorption feature in the middle corresponds to a dim feature in the left panel. It may be also associated with the bright EIT feature in the right panel of Figure \ref{fig6} indicated with the plus sign here.
}
\label{fig7}
\end{center}
\end{figure}

The left panel of Figure \ref{fig7} shows an H$\alpha$ image taken at 18:46:31 before the flare onset. It has many bright features, most of which have an EUV counterpart in the EIT images as shown for example in the left panel of Figure \ref{fig6}. The similarity of these images suggests that the EUV and H$\alpha$ features might be strongly coupled. However, the H$\alpha$ image is also quite different from the EUV image. In particular, there are many dark fibrils in the H$\alpha$ image, which are not evident in the EUV image. These cooler H$\alpha$ features should have a lower altitude than the warmer EUV features. Recently \citet{l12} studied the formation of H$\alpha$ line in the solar chromosphere and found that brighter H$\alpha$ features have lower average formation height. The bright H$\alpha$ features therefore should have even lower altitudes than the dark features.

The right panel of Figure \ref{fig7} shows the difference map between two H$\alpha$ images taken at 18:46:31 and 18:48:31 with the former treated as a preflare background. The overlaying contours show the 3-25 keV RHESSI image from 18:48:01 to 18:49:01, which is centered at the time of the latter H$\alpha$ image. While the north rim of the RHESSI arc appears to be well aligned with the north rim of the enhanced H$\alpha$ emission, no enhanced H$\alpha$ emission is seen along the south rim of the X-ray source. Since brighter H$\alpha$ features have lower average formation height, the enhanced H$\alpha$ emission is likely caused by reduction of H$\alpha$ opacity in the chromosphere by the flaring loop instead of emergence of new relatively cool H$\alpha$ emitting materials within the hot flare loop. One therefore may associate the absence of enhanced H$\alpha$ emission along the south rim of the X-ray loop with absorption features in the chromosphere with a linear size of about $10^9$ cm. Although the H$\alpha$ source extends further westward than the X-ray arc, all these emissions can be readily associated with a flare loop and there is no evident footpoint emission as seen in typical large two-ribbon flares.

\begin{figure}[ht]
\begin{center}
\includegraphics[width=15cm]{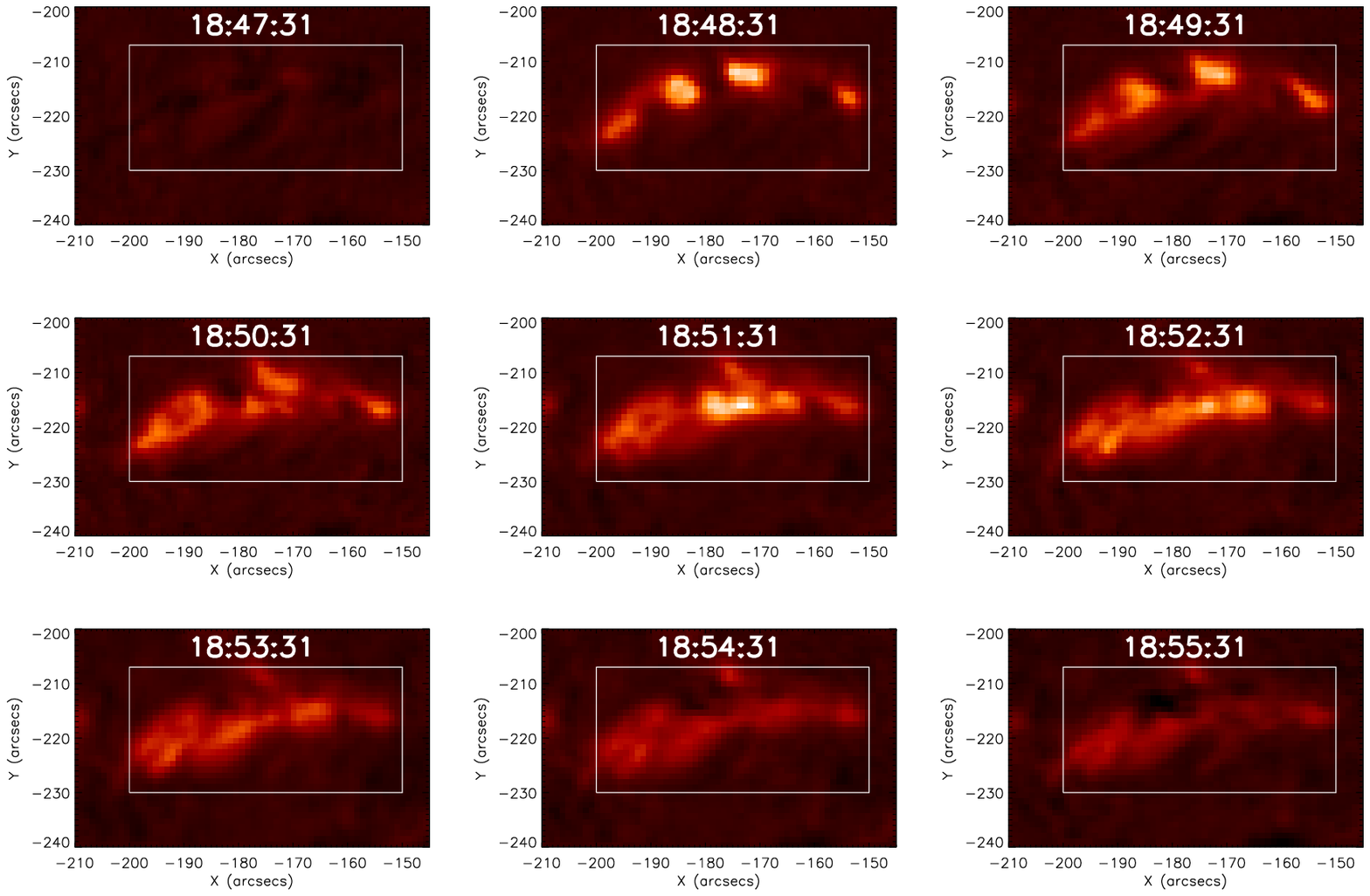}
\vspace{-0mm}
\caption{
Evolution of the flare loop in H$\alpha$. The H$\alpha$ image at 18:46:31 has been subtracted from each image to better illustrate emission associated with the flare. The color scale is the same for all frames to better demonstrate the brightness variations. The box is the same as that in the right panel of Figure \ref{fig3}.  The bright tips near the east, north, and west boundaries of the box appear to be stable.
}
\label{fig8}
\end{center}
\end{figure}

Figure \ref{fig8} shows the evolution of the H$\alpha$ loop. The pre-flare image at 18:46:31 in the left panel of Figure \ref{fig7} has been subtracted from all these images to illustrate the emission component associated with the flare.
At 18:47:31, when the RHESSI count rates from S2 just start to increase (Figs. \ref{fig1}, \ref{fig3}), there is no evidence of enhanced H$\alpha$. Enhanced H$\alpha$ emission is first detected one minute later after the HXR pulse. It is unfortunate that there are no H$\alpha$ observations during the HXR pulse. Both emission and absorption features evolve during the flare. In general, the brightest region moves south and absorption features to the south become thinner and thinner as the flare progresses. A dark absorption cloud forms to the north after 18:54:31. This absorbing material evolves significantly during the flare and may participate into the flare process.

The small size and rapid evolution of the absorbing material implies that it can not be very high in the corona. The altitude of the flare loop should be even lower so that the H$\alpha$ images of the flare loop can be affected by these materials. Otherwise a flare loop high in the corona needs to affect the low-lying H$\alpha$ absorption materials in a complicated manner to account for the correlated X-ray and H$\alpha$ images. The good correlation between H$\alpha$ and 195 \AA\ images discussed above also supports the view that the H$\alpha$ brightening during the flare is caused by heating of absorbing materials, which reduces the H$\alpha$ opacity. The flare loop therefore must lie very low in the solar corona or even in the chromosphere. It is interesting to note that the bright spot in the difference map of EUV emission in the right panel of Figure \ref{fig6} (indicated by the plus sign in the figure) is located at the dark region between the two bright H$\alpha$ spots in the middle, which may suggest interaction of the H$\alpha$ absorbing material with the EUV loop. The appearance of the H$\alpha$ images is strongly affected by absorption. Detailed spectroscopic observations of similar loops at H$\alpha$ will be able to determine the relative altitude of the loop and the absorption materials and the nature of the H$\alpha$ brightening.

\begin{figure}[ht]
\begin{center}
\includegraphics[width=15cm]{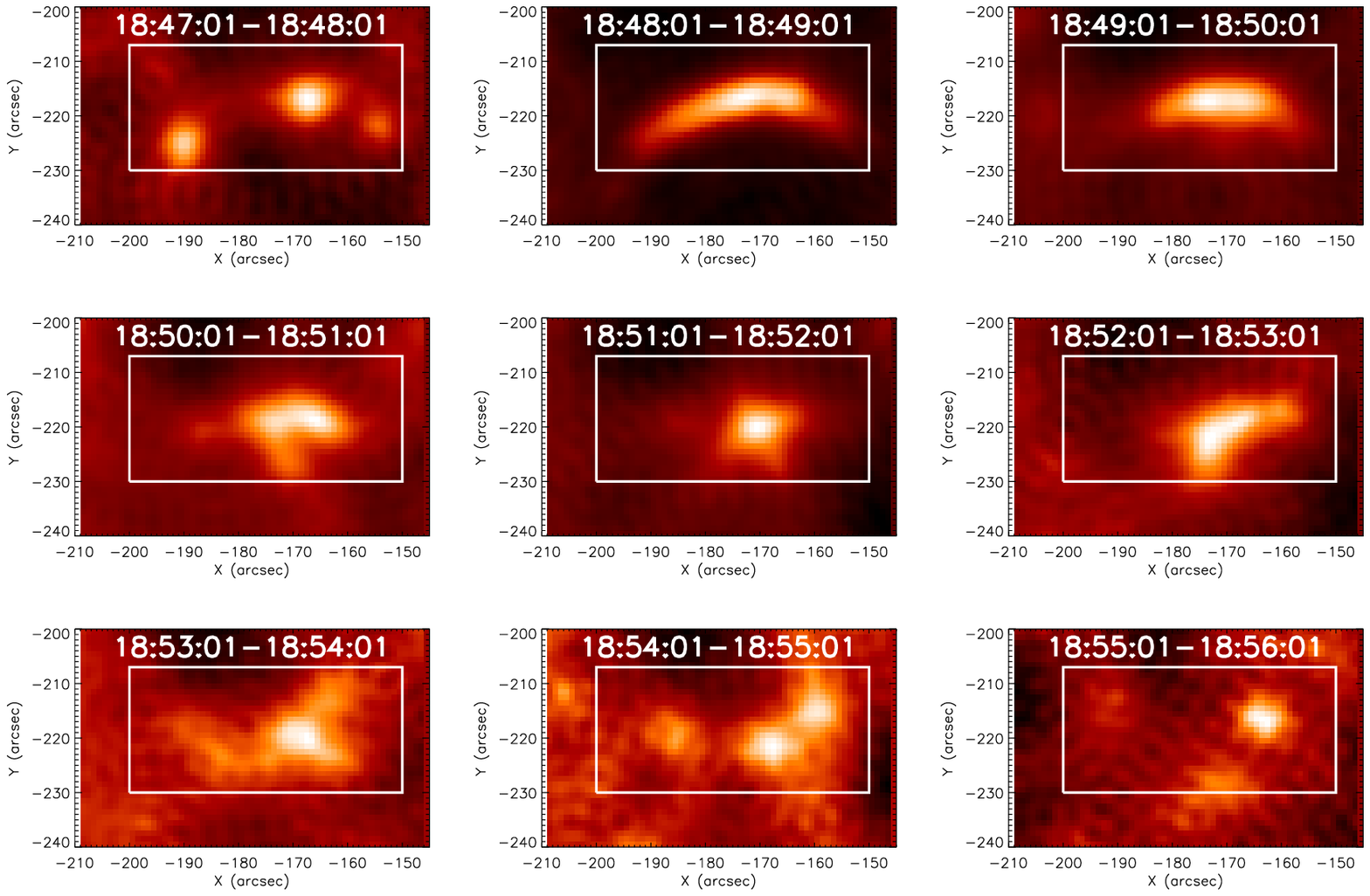}
\vspace{-0mm}
\caption{Evolution of the X-ray source observed with RHESSI. To compare with H$\alpha$ observations, images are made for every one minute interval centered on the times when the H$\alpha$ images are taken.
The front segment of all detectors except for detector 2 are used and the energy band is 3-25 keV. The box is the same as that in Figure \ref{fig8}. Since we are primarily interested in the evolution of the source structure, the color scale is chosen independently to optimize the appearance of each frame.
}
\label{fig9}
\end{center}
\end{figure}

Figure \ref{fig9} shows the evolution of the X-ray loop with a cadence of 1 minute to compare with the H$\alpha$ images in Figure \ref{fig8}.
Although both the X-ray and H$\alpha$ emission evidently come from the same flare region, fine structures of the X-ray are not well correlated with those of H$\alpha$ in Figure \ref{fig8}.
The arc structure in the X-ray can be seen only during the first three minutes. The X-ray source becomes irregular afterward, which is in contrast to the H$\alpha$ observations, where the source structure appears to evolve continuously.
Since X-ray emission is associated with the hottest plasmas and H$\alpha$ emission comes from relatively cold materials and is affected by absorption, differences between X-ray and H$\alpha$ images are expected, which is quite different from impulsive hard X-ray and H$\alpha$ emission caused by energetic electrons injected into chromospheric footpoints of classical flare loops.
 The appearance of three X-ray sources in the first time interval 18:47:01-18:48:01 is reminiscent of the classical flare model with one coronal source and two chromospheric footpoints. However, the X-ray source evolution, observations in other wavelengths, and the fact that all these sources are dominated by low energy photons suggest that they are all thermal sources along the coronal loop, which agrees with the energy dependent source structure shown in Figure \ref{fig4}.

\section{DISCUSSION AND CONCLUSIONS}
\label{dis}

In this paper we carried out a detailed analysis of multi-wavelength observations of a flare loop with distinct impulsive and gradual emission components. It is shown that all of the observed H$\alpha$ and X-ray emissions are associated with the coronal part of the loop with no evidence of emission from the footpoints and the impulsive component might be associated with superhot plasmas produced in turbulent-current sheets located in the loop \citep{lm93, s98, h07}, in contrast to the conventional view where the impulsive component is associated with nonthermal particles in the chromosphere. Although both an isothermal model and a power-law model give a comparable fit to the spatially integrated X-ray spectra in the impulsive phase, the lack of footpoint emission, the softness of the spectra with the power law model, and the complex X-ray and H$\alpha$ structure along the loop all favor a quasi-thermal interpretation of the impulsive component.

The flare loop studied here has very simple light curves, a very stable large scale structure, and appears to have a very low altitude with both emission and absorption features shown in the H$\alpha$ images.
The H$\alpha$ emission is correlated with the X-ray structure and aligned with the loop. The impulsive component is seen in both the HXR and H$\alpha$ light curves.
Detailed analysis of the evolution of the H$\alpha$ images suggests that the H$\alpha$ emission may be dictated by evolution of the absorption materials instead of emergence of new H$\alpha$ emitting plasmas within the loop.
Although the flare loop is not well covered by EUV observations, the absence of detectable 195 \AA\ emission at the HXR peak time is puzzling for the high temperature ($>20$ MK) inferred from RHESSI spectral fit. It could be caused by the delayed appearance of EUV emitting plasma in a flare loop \citep{r09}. Better observations of similar flare loops, especially spectroscopic observations in H$\alpha$, and detailed magnetohydrodynamic simulations of internal kink instability \citep{h07} may be helpful to advance our understanding of this flare loop.

Although RHESSI spectral analysis of the spatially integrated fluxes is consistent with an isothermal component, X-ray and H$\alpha$ images also reveal that two flares with comparable X-ray and H$\alpha$ fluxes occurred simultaneously from two distinct active regions several light-seconds apart. These results show that the RHESSI spectral analysis alone can not distinguish different emission components as revealed by the images and it is essential to combine spectroscopy with imaging and multi-wavelength observations to constrain the nature of the underlying physical processes. We were able to distinguish these two flares with X-ray and H$\alpha$ images and the corresponding light curves. While the flare loop behaves like an early impulsive flare triggered by an instantaneous burst with higher energy emission peaking at an earlier time \citep{s07}, the other flare has relatively complex structure and behaves like a gradual flare with gradually evolving fluxes in different wavelengths well correlated. The presence of a second flare limits our interpretation of the spatially integrated results, but we can still identify the impulsive component in H$\alpha$, soft X-ray, and hard X-ray light curves and reach the conclusions discussed above.

The flare loop studied here has a length exceeding $5\times 10^9$ cm. The flare magnitude however is very small with a GOES B class. It however shows several distinct characteristics: impulsive thermal emission, co-spatial H$\alpha$ and X-ray emission, lack of EUV emission at the HXR peak, relatively long and stable structure. The coincidence of this flare loop with another flare of comparable magnitude might be accidental, but nevertheless is also rare. The energetics of these flares is consistent with transient brightenings in active regions discovered by Yohkoh \citep{s92}. Such flare loops may occur frequently in contrast to what might be suggested by the unique characteristics present in this paper. Systematic study of similar flare loops is therefore warranted.

This study demonstrates clearly that solar flares are complex phenomena with each flare likely having some unique characteristics. Detailed study of flares with good observational coverage therefore is always warranted and might reveal features challenging the conventional wisdom, such as the impulsive thermal emission revealed here. Therefore while seeking common physical processes operating in solar flares, we should also pay due attention to the complexity of flares introduced by the complex magnetic environment of the solar active region, where all micro- and bigger flares originate \citep{c08}.


\acknowledgements
We thank the Big Bear Solar Observatory, New Jersey Institute of Technology for providing their H$\alpha$ observations and the RHESSI team for assistance with the RHESSI data analysis. SL thanks Hui Li, Hugh Hudson, Pengfei Chen, and Peter J. Cargill for helpful discussions.
This work is supported by the National Natural Science Foundation of China via the grant 11143007, 11173063, 11173064, 11233008, the EU's SOLAIRE
Research and Training Network at the University of Glasgow
(MTRN-CT-2006-035484) and STFC grant ST/1001808/1 and the EC-funded FP7 project HESPE (FP-2010-SPACE-1-263086).

{}

\end{document}